\documentclass{elsart}

\usepackage{natbib}

\usepackage{graphicx}
\usepackage{epstopdf}
\usepackage{enumerate}

\usepackage{amsmath}
\usepackage{amssymb}

\usepackage{enumerate}
\usepackage{pslatex}

\newtheorem{theo}{Theorem}
\newtheorem{coro}{Corollary}

\newcommand{\EE}{\mathbb{E}}

\begin{document}

\begin{frontmatter}



\title{The Shared Reward Dilemma}


\author[gisc]{J.~A.~Cuesta},
\author[estadistica,bolivar]{R.~Jim\'enez\corauthref{corresp}}
\author[economia]{H.~Lugo}
\author[gisc,imdea,bifi]{A.~S\'anchez}

\address[gisc]{Grupo Interdisciplinar de Sistemas Complejos (GISC),
Departamento de Matem\'aticas, Universidad Carlos III de Madrid, Legan\'es, Spain}
\address[estadistica]{Departamento de Estad\'{\i}stica, Universidad Carlos
III de Madrid, Getafe, Spain}
\address[bolivar]{Departamento de C\'omputo Cient\'{\i}fico y Estad\'{\i}stica,
Universidad Sim\'on Bol\'{\i}var, Caracas, Venezuela}
\address[economia]{Departamento de Econom\'{\i}a, Universidad Carlos III de
Madrid, Getafe, Spain}
\address[imdea]{IMDEA Matem\'aticas, Madrid, Spain}
\address[bifi]{Instituto de Biocomputaci\'on y F\'{i}sica de Sistemas Complejos,
Universidad de Zaragoza, Zaragoza, Spain}

\corauth[corresp]{\emph{Corresponding author:} Departamento de Estad\'istica,
Universidad Carlos III de Madrid,  C/ Madrid, 126 - 28903 Getafe (Madrid) Spain.
Email: raul.jimenez@uc3m.es.}

\begin{abstract}
One of the most direct human mechanisms of promoting cooperation
is rewarding it.
We study the effect of sharing a reward among cooperators in
the most stringent form of social dilemma, namely the Prisoner's Dilemma.
Specifically, for a group of players that collect payoffs by playing a
pairwise Prisoner's Dilemma 
game with their partners, we consider an external entity that distributes
a fixed reward equally among all cooperators. Thus, individuals
confront a new dilemma: on the one hand, they may be inclined to choose the
shared reward despite the possibility of being exploited by defectors; on the other
hand, if too many players do that, cooperators will obtain a poor reward and defectors will
outperform them. 
By appropriately tuning the amount to be shared a vast variety of scenarios
arises,
including traditional ones in the study of cooperation as well as
more complex situations where unexpected behavior can occur.
We provide a complete classification of the equilibria of the $n$-player game
as well as of its evolutionary dynamics.
\end{abstract}


\end{frontmatter}


\section{Introduction}
\label{sec:intro}

Selfish behavior seems to be one of the consequences of evolutionary dynamics. Genes, organisms,
generic entities acting in their own benefit do better in a struggle for reproductive
(understood in a wide sense) success and are selected in the long term. In spite of this
general trend, we find in every evolutionary context (be it biological, sociological,
economic, etc.) many instances in which cooperative behaviors are evolutionarily
successful. The explanation of this puzzle has developed into an active line of research,
and providing a complete answer to it is one of the big open problems of XXI century
\citep{pennisi}.
Many mechanisms have been identified as responsible for these cooperative associations.
Among them we find kinship \citep{hamiltona,hamiltonb}, reciprocity \citep{axelrod:1981},
reputation gain \citep{nowaksigmund:1998}, and others \citep{axelrod:1984,nowak:2006}. 
One of the most interesting mechanisms of this kind that has been identified is 
altruistic punishment and rewarding \citep{sigmund:2001} or voluntary participation
\citep{hauert:2007}. Through this mechanism
social groups that are engaged in social dilemmas, such as the one represented
by the Public
Goods game, can overcome the well-known tragedy of the commons \citep{hardin:1968}.

The rewarding mechanisms just mentioned are of the bottom-up type, i.e., they arise
at the individual level and lead to cooperation at the group level. 
However, in ecological and social contexts, there are several levels of 
organization which make possible top-down approaches. For instance, 
parents, educators, governments and other institutions promote prosocial 
behavior by rewarding individuals in different manners (prizes, incentives, tax
deductions, etc.). In biological or ecological contexts, some species reward
symbionts that cooperate at the required level by providing them with more 
resources (see
\citet{kiers:2003} and references therein).
 Companies also use similar mechanisms in their own 
benefit to induce customers to supply useful information about consumption 
habits or social networks \citep{iribarren}. Finally, another instance of
top-down rewarding can be found in 
team formation 
of animal societies \citep{anderson}, e.g. in cooperative hunting
\citep{packer}.

Top-down rewarding mechanisms can be generically implemented in two different
ways. The simplest one is to provide a fixed benefit to every cooperator. 
In terms of game theory, this is tantamount to shifting the payoff matrix
by a constant added to entries related to cooperation. Thus, for instance,
if one starts off with a Prisoner's Dilemma (PD) to model the baseline social 
behavior, introducing such a reward transforms the dilemma into another one,
either Snowdrift \citep{maynard-price:1973,sugden:1986}
or Stag Hunt \citep{skyrms:2003}, or even suppresses completely the 
dilemma, changing it into a Harmony game \citep{licht:1999}.
A second, more subtle mechanism is to distribute a fixed amount between all
cooperators in the population. In this case, the original PD becomes
a new dilemma, because there is a clear incentive to 
cooperate but if there are too many cooperators the incentive disappears
and hence defecting pays. This is reminiscent of the Minority game 
paradigm \citep{moro} and, in fact, it may be seen as an alternative form 
of describing situations in which being in the minority (understood in 
a lax sense) is the best option. We will refer to this situation as the 
{\em shared reward dilemma}. 

In this work we study the shared reward dilemma by considering
an interaction group of $n$ individuals.
In order to understand it in the most stringent form of social
dilemma, interaction among individuals follows the PD
(see \citet{doebeli-hauert:2005} for a review).
Thus, we introduce a game in which payoffs can be obtained from two sources:
first, all players collect payoffs by playing a $n$-player generalization
of the PD game with their partners \citep{hauert:2003}, and
second, players who have chosen to cooperate share  an extra payoff
coming from a  pool. In the next
section we analyze in detail the $n$-player  game.
Situations in which multiple interior equilibria occur are completely determined,
as well as the parametric settings in which equilibria increase, decrease or
jump discontinuously with the reward.
In Section~\ref{sec:evolution} we analyze the evolutionary stability 
of the equilibria discussed in Section~\ref{sec:thegame} and provide the different
asymptotic scenarios of cooperation
according to the replicator dynamics.
Section~\ref{sec:conclusions} summarizes our conclusions and presents 
some future prospects. Appendix~\ref{sec:app-sym} contains
the main mathematical results on which the discussions of previous sections
rest: a theorem and a 
corollary that provide closed formulae for the symmetric Nash equilibria in terms of the
reward for finite and large number of players, respectively. To complete our analysis,
we present in
Appendix~\ref{sec:app-asym}
a theorem which characterizes all asymmetric Nash equilibria in pure
strategies of the game. 

\section{The shared reward dilemma}
\label{sec:thegame}

Consider an assembly of $n$ players, each of whom can choose one out of two actions:
cooperate (C)  or defect (D) with the rest of the $n-1$ players
in an one-shot game (i.e., all player's actions are simultaneously performed).
Players collect  payoffs according to a PD game
from every one of the $n-1$ opponents.
In addition, players who have chosen to cooperate obtain an extra payoff
coming from a fixed reward $\rho$,
provided by an external source,
that is evenly distributed among all cooperators.

To provide the strategic form of this game we introduce some notation.
Let $k$ be the number of cooperators in the group. 
Payoffs of pairwise interactions
are denoted by the standard parameters of the PD game: a defector that exploits
a cooperator obtains the temptation $T$, but when she faces up another defector
she receives the punishment $P$; instead, the payoff for a cooperator
meeting another cooperator is 
the reward $R$ (not to be confused with $\rho$,
the reward to be shared  that we propose in this
work), but obtains the sucker's payoff $S$ when she confronts a defector. For the
game to be a PD, the payoff  must be ordered according to $T>R>P>S$. 
Since the game is symmetric, in the sense that the payoff to a particular
player is independent of her label and only depends on her actions,
the total payoff of an arbitrary player
is given by
\begin{equation} 
U = \begin{cases}
(k-1) R + (n-k)S + \displaystyle{\frac{\rho}{k}}, &\mbox{if she
cooperates,} \\
k T + (n-1-k) P, & \mbox{if she defects}.
\end{cases}
\label{pagototal}
\end{equation}
The remaining
of this section is devoted to study the Nash equilibria  of this game.

Let us begin with the symmetric Nash equilibria in pure strategies,
which can be easily obtained from (\ref{pagototal}).
Full cooperation is an equilibrium if 
no player increases her payoff by defecting unilaterally,
that is, if and only if $T(n-1) \leq (n-1)R + \rho/n$.
Similarly, full defection is an equilibrium if 
no player increases her payoff by cooperating unilaterally, i.e., 
if and only if $(n-1)S + \rho \leq (n-1)P$.
The former constraint on $\rho$ suggests a normalization of the shared
reward, namely
\begin{equation}
\delta=\frac{\rho}{n(n-1)(T-R)},
\end{equation}
which will henceforth be referred to as \emph{scaled reward}.
With this parameter, the condition for full cooperation to be a Nash
equilibrium is simply $\delta\ge 1$.
As for the second constraint, if we introduce a new parameter,
the \emph{defection ratio}
\begin{equation}
\zeta=\frac{T-R}{P-S},
\end{equation}
the condition for full defection to be a Nash equilibrium is
$\delta\le 1/n\zeta$. All the analysis of the game can be performed
solely in terms of these two parameters instead of the five parameters that
originally define the game. 
As we have shown, the scaled reward is the ratio
between the actual reward and the reward needed for full cooperation 
to be a Nash equilibrium; as for the
defection ratio, it compares, in a pairwise interaction,
the excess of payoff a defector gets over a cooperator
when both confront a cooperator, with that when both face up a defector.

Note that both full defection and full
cooperation will coexist if and only if $1\le\delta\le 1/n\zeta$.
Clearly, no reward meets this condition unless $\zeta\leq 1/n$.
Thus we see that, by increasing the reward, the
symmetric Nash equilibrium in pure strategies
changes from full defection to full cooperation, and in between
these two extremes there may be either coexistence or absence of
both equilibria, depending on whether $\zeta$ is smaller or larger than $1/n$,
respectively.

The space of symmetric mixed strategies Nash equilibria consists of all $0\leq q \leq 1$
such that a player cooperates with probability $q$ and defects with probability $1-q$.
The expected total payoffs of an arbitrary cooperator and of an arbitrary defector
when the rest of the players play an equilibrium $q$, are given by 
\begin{eqnarray}
\label{fitness1}
f_C(q) &=& \EE[U|\mbox{she cooperates}] = (n-1)qR + (n-1)(1-q)S + \rho \mu_{n-1}(q), \\
\label{fitness2}
f_D(q) &=& \EE[U|\mbox{she defects}] = (n-1)qT + (n-1)(1-q)P, 
\end{eqnarray}
where $\mu_m(q) = \EE[(S_m+1)^{-1}]$, $S_m$ being a binomial random variable which
is the sum of $m$ i.i.d.\ Bernoulli's random variables with mean $q$.
As has been observed by \citet{chao:1972}, $\mu_{m}(q)$ has the expression 
\begin{equation} 
\label{mu1}
\mu_{m}(q) = \begin{cases}
1, & \mbox{for $q=0$,} \\
\displaystyle{\frac{1-(1-q)^{m+1}}{(m+1)q}},  & \mbox{for $0<q\le 1$.}
\end{cases}
\end{equation}
Symmetric Nash equilibria in completely
mixed strategies can be computed by solving $f_C(q) = f_D(q)$.
To do that, it is convenient to distinguish when there are  more than two players
and when there are just two players involved.
The latter case is particularly simple because it reproduces the major binary games
used in the study of cooperation. The payoff matrix  \citep{Gintis:2000} of this binary game can
be easily obtained from (\ref{pagototal}) by setting $n=2$, and it is shown in
Table~\ref{SR}. Thus, depending on $\rho$, the game becomes a:
\begin{enumerate}[(i)]
\item Prisoner's Dilemma, if $T>R+\rho/2$ and $P>S+\rho$;
\item Snowdrift, if $T>R+\rho/2$ and $P<S+\rho$; 
\item Stag-hunt, if $T<R+\rho/2$ and $P>S+\rho$; 
\item Harmony, if $T<R+\rho/2$ and $P<S+\rho$. 
\end{enumerate}

\begin{table}[!ht] 
\begin{center}
\begin{tabular}{c|c|c}
              & C & D \\ \hline
C    &  $R +\rho/2$    & $S+\rho$         \\ \hline
D    &  $T$                        & $P$ \\  \hline
\end{tabular}
\end{center}
\caption[]{Payoff matrix for the binary case of the shared reward dilemma.}
\label{SR}
\end{table}

The Nash equilibria of these games are well known. 
Thus, the Snowdrift game has two asymmetric Nash
equilibria in pure strategies, \{(C,D), (D,C)\},
while the Stag-hunt game has two symmetric Nash equilibria,  \{(C,C), (D, D)\}.
Both games have a unique Nash equilibrium in mixed strategies $q\in(0,1)$.
Otherwise, the Prisoner's Dilemma and the Harmony game have just one Nash
equilibrium (both players  defecting and both cooperating, respectively).

In terms of $\delta$ and $\zeta$,
the above conditions (i)--(iv) can be rephrased as
\begin{enumerate}[(i')]
\item Prisoner's dilemma if $\delta<\min(1,1/2\zeta)$;
\item Snowdrift if $1/2\zeta<\delta<1$;
\item Stag-hunt if $1<\delta<1/2\zeta$;
\item Harmony if $\delta>\max(1,1/2\zeta)$.
\end{enumerate}

In general, our results permit to characterize
the changes in the structure of equilibria by
varying $\delta$ and fixing $\zeta$. Therefore, we
can study the effect of rising the reward. In order
to illustrate our approach, consider once more the binary game.
Upon increasing $\delta$ the game changes from Prisoner's dilemma to Harmony.
For $\zeta=1/2$ this change occurs directly when $\delta$ crosses at 1,
but depending on whether $\zeta>1/2$ or $\zeta<1/2$, the
change occurs via Snowdrift or via Stag-hunt, respectively.

\begin{figure}
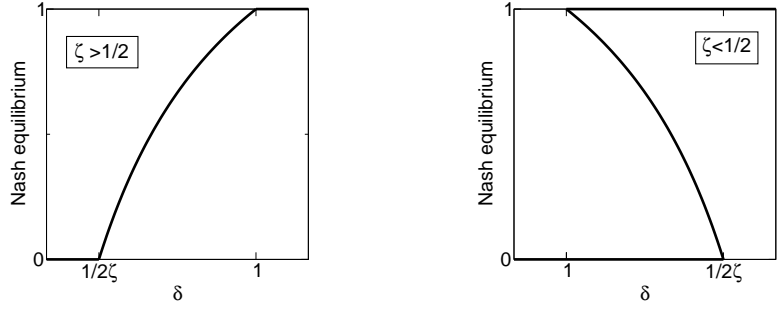

\begin{center}
\includegraphics[width=40mm]{twoplayers-1.eps} \quad\qquad\qquad
\includegraphics[width=40mm]{twoplayers-2.eps} 
\end{center}
\caption[]{Symmetric Nash equilibria of the binary game as a function of the
scaled reward $\delta$ for the two types of possible behavior,
$\zeta>1/2$ (left) and $\zeta<1/2$ (right).}
\label{fig:twoplayers}
\end{figure}

Taking $n=2$ in (\ref{fitness1}) and (\ref{fitness2}) (hence $\mu_1(q)=1-q/2$)
and solving $f_C(q) = f_D(q)$ we obtain a unique Nash equilibrium in mixed strategies
$0<q<1$ given by
\begin{equation}
q=\frac{1-2\delta\zeta}{1-(1+\delta)\zeta}.
\end{equation}
If $\zeta>1/2$ (respectively $\zeta<1/2$) $q$ is a continuous increasing (respectively
decreasing) function of $\delta$.
Figure~\ref{fig:twoplayers} illustrates these two scenarios as well as
the parametric conditions for the existence and coexistence of equilibria in pure strategies.
When $\delta$ lies in between 1 and $1/2\zeta$, there is uncertainty as to the
strategy that players will choose: for $\zeta>1/2$, because no symmetric Nash equilibrium
in pure strategies exists when $1/2\zeta<\delta<1$; for $\zeta<1/2$, because there is
coexistence of both full cooperation and full defection in the range $1<\delta<1/2\zeta$.
In the former case the mixed strategies Nash equilibrium that fills the gap has
the expected behavior: the probability of cooperating increases with the reward;
however, in the latter case the behavior of this Nash equilibrium is counterintuitive,
as the probability of cooperating decreases with the reward.
This phenomenon can be explained in the framework of evolutionary dynamics,
where the binary game models pairwise interactions between
individuals of a large population.
In this context, it is well known that, under
the replicator dynamics, the equilibrium in mixed strategies of
the Stag-hunt game is unstable and separates the basins of attraction
of the two equilibria in pure strategies
 (full defection and full cooperation).
We will come back to this issue in Section~\ref{sec:evolution} in a more general setting,
where we study in detail the replicator dynamics  by considering
interactions in groups of $n$ individuals.

\begin{figure}
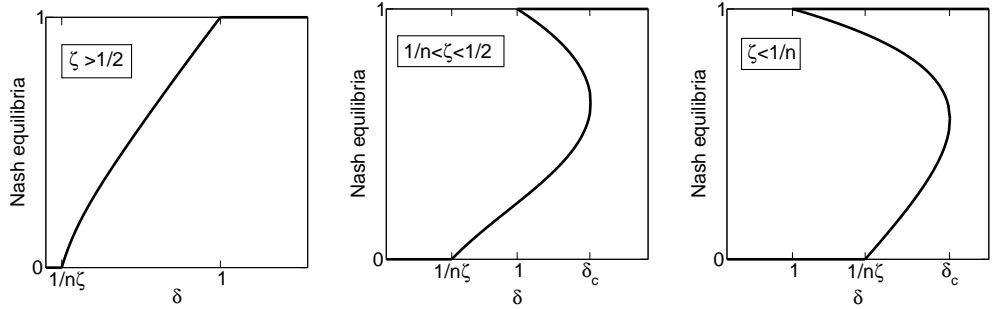

\begin{center}
\includegraphics[width=40mm,clip=]{case1.eps} \quad
\includegraphics[width=40mm,clip=]{case2.eps} \quad
\includegraphics[width=40mm,clip=]{case3.eps}
\end{center}
\caption[]{Symmetric Nash equilibria of the $n$-player game ($n\geq3$) as a
function of $\delta$ for the three types of possible behavior,
$\zeta>1/2$ (left) and $1/n<\zeta<1/2$ (middle) and $\zeta<1/n$ (right).}
\label{fig:cases}
\end{figure}

Let us now analyze the case $n \ge 3$. Notice that $\mu_{n-1}(q)$ defined in
(\ref{mu1}) is now a nonlinear function of $q$ and thus there can be more than one
solution of $f_C(q) = f_D(q)$. However, as such solutions are obtained as
the intersection points of a straight line with a strictly convex function,
there can be up to two equilibria in the open interval $(0,1)$. 
As is proven in Theorem~\ref{th1} of Appendix~\ref{sec:app-sym},
the number of equilibria depends only on
the values of $\delta$ and $\zeta$. Moreover,
the changes on the structure of equilibria when $\delta$ increases
correspond to three possible scenarios, determined by
$\zeta<1/n$,  $1/n\leq \zeta <1/2$ and $\zeta\geq 1/2$.
(Notice that for $n=2$ the middle case is empty, and the other two
cases correspond to those discussed above.)
Figure~\ref{fig:cases} depicts the typical structure of equilibria
for these three cases.
 
For the case $\zeta\geq1/2$, Theorem~\ref{th1} shows
that there exists a unique symmetric Nash equilibrium which
is a continuous increasing function of $\delta$. It is strictly
increasing within $[1/n\zeta, 1]$ from full defection at $\delta=1/n\zeta$
to full cooperation at $\delta=1$, and constant outside the interval.
However, when $\zeta<1/2$ we have two nontrivial, different scenarios.
One feature common to both of them is the existence of a range of rewards,
namely $\max\{1,1/n\zeta\}<\delta< \delta_c$, for
which two symmetric equilibria in mixed strategies coexist.
One of these equilibria increases and the other 
decreases when the reward increases within this range.
At the critical value $\delta_c$ these equilibria collapse and a further
increase in $\delta$ yields a discontinuous jump from a Nash equilibrium with
$q<1$ to full cooperation. An upper bound for $\delta_c$ is provided in
Theorem~\ref{th1}. The fundamental difference between the cases $\zeta<1/n$
and $1/n<\zeta<1/2$ arises in the region $\min\{1,1/n\zeta\}<\delta<
\max\{1,1/n\zeta\}$, where there exists a unique equilibrium $0<q<1$: for
$1/n< \zeta<1/2$ we see that $q$ increases with $\delta$, while for
$\zeta<1/n$, we see that $q$ decreases with $\delta$, exhibiting the
same counterintuitive behavior reported for the binary case.

A case of particular importance is $\zeta=1$,
because it reproduces the cost/benefit parametrization of the PD game, by
letting $T=b$, $R=b-c$, $P=0$ and $S=-c$, with $b>c>0$.
For this popular framework, suitable for biological applications,
our result shows that the equilibrium of the shared reward dilemma
only depends on the fixed amount $\rho$ to be shared by the cooperators
and on the cost $c$ of cooperation, but it is independent of the benefit $b$.
An analogous result is observed in a spatial
evolutionary version of the shared reward dilemma \citep{auto1}.

When the number of players $n\to\infty$, we provide a simplified
asymptotic version of Theorem~\ref{th1}, in Corollary~\ref{cor1}
of the Appendix~\ref{sec:app-sym}.
As in this limit the threshold $1/n\zeta \rightarrow~0$,
the third
of the three cases shown in Figure~\ref{fig:cases} disappears.
Notice that in order to get $0<\delta<\infty$ in the $n\to\infty$ limit,
we have to scale the reward with the number of interactions in the game, $n(n-1)$.
The reason is that
the payoffs collected per player from their pairwise interactions, in the
first step of the game,  are $O(n)$, therefore
the reward per player must be of the same order
to produce an effect.  This makes $\rho=O(n^2)$.
In that case,
the shapes of the first two cases in Figure~\ref{fig:cases} 
are preserved, with a shift of the 
threshold $1/n\zeta$
to 0 (full defection is an equilibrium if and only if  $\rho=o(n^2))$.
The critical value of the scaled reward, $\delta_c$,
at which the equilibrium jumps discontinuously from a value $q<1$ to
full cooperation when $\zeta<1/2$, can be exactly computed in the asymptotic case $n\to\infty$.
As it is proved in Corollary~\ref{cor1}, $\delta_c = 1/4\zeta(1-\zeta)$.

The limit case
$\zeta\to+\infty$ (equivalent to $P\rightarrow S^+$) has also received
special attention in the analysis of PD games on complex networks
\citep{nowak:2000, maxy:2005}. Our results show (c.f.\ eq.~(\ref{algeq}))
that a well defined mixed Nash equilibrium exists for $0<\delta<1$
which monotonically increases with $\delta$ from 0 to 1, reaching full
cooperation for $\delta\ge 1$. In the $n\to\infty$ limit, using
Corollary~\ref{cor1},
we can obtain an estimate for the equilibrium when $P\rightarrow S^+$,
namely the smallest value between $\sqrt{\delta}$ and 1.

Asymmetric Nash equilibria in pure strategies, in which part of
the players in the group cooperate
and the rest defect, can also be found for this game. 
For an interval of rewards starting at $1/n\zeta$
(the maximum reward for which full defection is a Nash
equilibrium) there exist asymmetric equilibria with $k$ cooperators and
$n-k$ defectors. The value of $k$ increases stepwise, starting from $k=1$,
at reward values $1/n\zeta=\delta_1<\delta_2
<\ldots$ (see eq.~(\ref{steps})), with equilibria with $k$ and $k+1$
cooperators coexisting precisely 
and only at the separating values $\delta_k$. For instance, 
upon increasing $\delta$ above $1/n\zeta$, the full defection equilibrium is
replaced by one with a single cooperator and $n-1$ defectors. 
In turn, this is the only Nash
equilibrium in pure strategies up $\delta_2$, where it is replaced
by another equilibrium with two cooperators and $n-2$ defectors.
The maximum number of cooperators in asymmetric equilibria is
$n-1$ if $\zeta\geq 1/2$, or else the largest
integer $k \leq (n-1)/2(1-\zeta)$ if $\zeta<1/2$.
In order to complete the analysis of the static game, a full characterization
of these equilibria is given by Theorem~\ref{th2} of Appendix~\ref{sec:app-asym}.
There is a particular aspect of them which we would like to call attention
upon: the fraction of cooperators in the asymmetric Nash equilibria approaches
either the unique or the lowest mixed strategies Nash equilibrium $0<q<1$ in the limit
$n\to\infty$. As we will see in Section~\ref{sec:evolution}, for the study
of the replicator dynamics based on the shared reward dilemma,
only the knowledge of symmetric Nash equilibria is necessary.

\section{Evolutionary dynamics}
\label{sec:evolution}

In population dynamics, the evolution of cooperation can be modeled in
several ways. According to the replicator dynamics \citep{hofbauer-sigmund:1998},
the dynamics in infinitely large populations is described by
\begin{equation}
\frac{dx}{dt}=x(1-x)\big[f_C(x)-f_D(x)\big],
\label{eq:replicator}
\end{equation}
$x(t)$ being the fraction of cooperators at time $t$ and $f_C(x)$ and
$f_D(x)$ the average fitness (which is the evolutionary counterpart of the
concept of payoff) of cooperators and defectors in the
population, respectively.
In this paper we consider the approach presented by \citet{hauert:2006}
to study replicator dynamics based on interaction groups of individuals.
The standard setup to obtain the replicator equation is to assume a large
population of individuals who randomly select partners to play a two-person
game.
In this alternative approach, players select groups of $n-1$ individuals and play an
$n$-person game instead. This is an appropriate approach to study the evolutionary
behavior of populations interacting through Public Goods games
\citep{hauert:2006}, and it is also suitable to study the
evolutionary behavior of the shared reward dilemma.

If the population is well-mixed, the number of cooperators at time $t$ in an interaction
group of $n$ individuals is a binomial random variable with mean $nx(t)$.
Therefore, the average fitnesses at time $t$ 
are given by formulae (\ref{fitness1}) and (\ref{fitness2}) with $q= x(t)$.
Inserting these formulae in (\ref{eq:replicator})
we model the evolution of cooperation when a reward $\rho$ is available for
each interaction group.

\begin{figure}
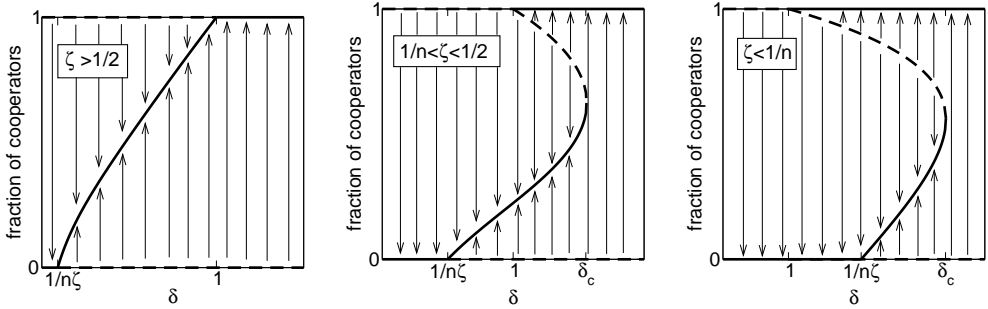

\begin{center}
\includegraphics[width=40mm,clip=]{case1-evol.eps} \quad
\includegraphics[width=40mm,clip=]{case2-evol.eps} \quad
\includegraphics[width=40mm,clip=]{case3-evol.eps}
\end{center}
\caption[]{Equilibria
of the replicator equation (\ref{eq:replicator}). Solid lines represent the asymptotically
stable fixed points, while dashed lines represent the unstable ones.}
\label{fig:stability}
\end{figure}

It is clear that $x=0$ and $x=1$ are always fixed points of the replicator equation
(\ref{eq:replicator}), but there will be further fixed points at the solutions of
$f_C(x^*)=f_D(x^*)$ in the open interval $(0,1)$. All of them are the symmetric Nash
equilibria discussed in previous section. By the \emph{folk theorem} of
evolutionary game theory \citep{Cr03}, the asymptotic stability of these fixed points
will depend on the sign of $f_C(x)-f_D(x)$. For example, if it is always positive,
$x=0$ is unstable whereas $x=1$ is stable,
and if it is always negative it is the other way around. The situation is 
different if
$f_C(x)-f_D(x)$ changes sign in the interval $(0,1)$. By Theorem~\ref{th1} (see
Appendix \ref{sec:app-sym}),
we can determine how many roots (none, one or two)
has  $f_C(x)-f_D(x)$ in the open interval $(0,1)$.
On the other hand, since $f_C(0)-f_D(0)=n(n-1)(T-R)(\delta-1/n\zeta)$,
then $x=0$ is stable if $\delta<1/n\zeta$ and it is unstable otherwise.
Thus, we will find the following stability patterns, depending on the number
of roots of (\ref{algeq}) in the interval $(0,1)$:
\begin{enumerate}[(I)]
\item if $\delta<1/n\zeta$ (in this case there is either none or just one root),
\begin{enumerate}[(a)]
\item if there are no roots, $x=0$ is a stable and $x=1$ an unstable
fixed point;
\item if there is one root $0<x_1<1$, then $x=0$ is a stable, $x_1$
an unstable and $x=1$ a stable fixed point,
with $x_1$ separating the basins of attraction of $x=0$ and $x=1$;
\end{enumerate}
\item if $\delta>1/n\zeta$,
\begin{enumerate}[(a)]
\item if there are no roots, $x=0$ is an unstable and $x=1$ a stable
fixed point;
\item if there is one root $0<x_1<1$, then $x=0$ is an unstable,
$x_1$ is a stable and $x=1$ an unstable fixed point;
\item if there are two roots $0<x_1<x_2<1$, then $x=0$ is an unstable,
$x_1$ a stable, $x_2$ an unstable and $x=1$ a stable fixed point,
and $x_2$ separates the basins of attraction of $x_1$ and $x=1$.
\end{enumerate}
\end{enumerate}
All these situations are illustrated in Figure~\ref{fig:stability}.
Obviously the structure of fixed points of the replicator equation is the
same as that of the symmetric Nash equilibria described in the previous section.
The only difference is that now $x=0$ and $x=1$ are always fixed points.
What is really new is the stability patterns induced by the dynamics.
These patterns are shown in Figure~\ref{fig:stability} through flux
lines which indicate the direction in which the dynamics approaches
the stable equilibria. It is worth noticing that 
for the two cases with $\zeta<1/2$ (middle and right panels of Figure~\ref{fig:stability})
there is a critical value of the reward, $\delta_c$,
at which, starting from a zero fraction of cooperators, the 
asymptotic cooperation level jumps discontinuously from a value
$q<1$ to full cooperation. 
In both of them there is also a region of $\delta$ in which, depending on the
initial fraction of cooperators, the outcome may be full cooperation or a smaller
fraction of cooperators. This smaller fraction outcome may even be $0$ in the case in
which $\zeta<1/n$. An important consequence is that, $x=0$ being unstable for any
$\delta>1/n\zeta$, for a suitable reward, a single mutant in an interaction group
of defectors will spread cooperation in the population.

\begin{figure}
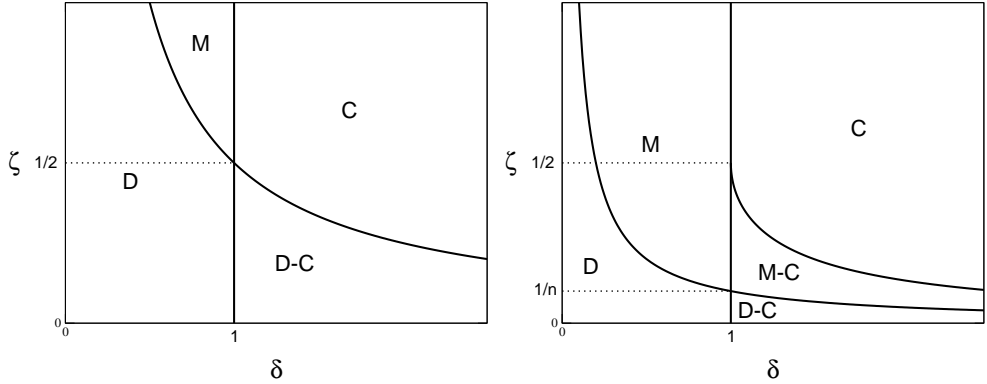

\begin{center}
\includegraphics[width=65mm,clip=]{diagram-2p.eps} 
\includegraphics[width=65mm,clip=]{diagram.eps} 
\end{center}
\caption[]{Diagrams sketching the different dynamical regimes for
$n = 2$ (left) and $n\geq 3$ (right) in terms of the two parameters $\zeta$
(defection ratio) and $\delta$ (scaled reward). Symbols stand for full defection
(D), full cooperation (C), co-existence of defectors and cooperators in a
mixed equilibrium (M), and existence of two stable
equilibria, either full defection/full cooperation (D-C) or mixed equilibrium/full
cooperation (M-C), each of which is reached depending on the fraction of cooperators
in the initial population. The curve marking the upper bound for the D and D-C regions
in both figures is given by $\zeta=1/n\delta$.
As $n\to\infty$ this curve moves towards the lower-left corner,
thus shrinking these two regions, which disappear in the strict limit.
The other curve of the right figure corresponds to the value of $\delta$
at which the two mixed equilibria which are found for $\zeta<1/2$ coalesce
($\delta_c$; see text).}
\label{fig:diagrams}
\end{figure}

To complete our analysis, we summarize the different dynamical regimes
that can be obtained, by varying $\delta$ and $\zeta$,
in Fig.~\ref{fig:diagrams}. These diagrams
illustrate the transitions between the different
evolutionary outcomes:
full defection, coexistence of cooperators and defectors,
bi-stability ---where full defection or full cooperation can be reached,
depending on the initial population---,
full cooperation, and ---only for $n\geq 3$ players---
bi-stability between a mixed population and full cooperation.

\section{Conclusions}
\label{sec:conclusions}

In this paper we have studied the effect of rewarding cooperation in a strict social
dilemma through the distribution of a fixed amount
among all cooperative individuals. By adding this payment to the standard payoffs
of the Prisoner's Dilemma, cooperators and defectors in an interaction group
confront a dilemma: on the one hand, individuals may be inclined to choose for
shared reward despite the possibility of being exploited by defectors; on the other
hand, if too many players do that, cooperators will obtain a poor reward and defectors will
outperform them. In the simplest case with only two players, we recover the
traditional binary games for the study of cooperation where the social dilemma is
relaxed: stag hunt and snowdrift.

Although intuition suggests that in this game there should be a threshold
value of the reward
above which cooperation increases monotonically up to reaching
saturation, the game exhibits more complex situations. The equilibrium 
structure has been characterized for the static game as well as for an
evolutionary version of the game based on the replicator dynamics. For a wide 
range of parameters, scenarios with multiple
interior equilibrium points are obtained, featuring critical values of the
reward at which cooperation jumps discontinuously. Also, counterintuitive behavior
where cooperation decreases as the reward increases may be observed. On the
other hand, the replicator dynamics provides additional 
stability criteria for these equilibria. In the
light of the stability patterns that arise, counterintuitive
equilibria in the static game, exhibiting a decrease of cooperation upon
increasing reward, turn out to be unstable equilibria of the dynamics
separating
basins of attraction of other stable equilibria. As a consequence, a most
relevant conclusion is that for many choices of the game parameters and 
initial conditions, the equilibrium with lower value of the cooperation
level is dynamically selected instead of the full cooperation one. 

The results presented in this paper allow for a complete characterization 
of the shared reward dilemma in the following terms. 
Cooperation does not appear until the reward increases above the
threshold $\delta=\min\{1,1/n\zeta\}$. Interestingly, 
for $\delta>1/n\zeta$, even a single cooperator can
spread cooperation in the population, the more the larger the reward. This is 
an important point supporting the effectiveness of the reward mechanism for
promoting the emergence of cooperation \citep{auto1}. Subsequently, for
$\zeta\ge 1/2$ the fraction of cooperators increases monotonically until full
cooperation is reached for $\delta=1$. However, and quite unexpectedly, 
for $\zeta<1/2$ an interesting 
phenomenon is observed: starting with a single cooperator,
full invasion of the population only takes place when the scaled reward $\delta>
\delta_c$, for some $\delta_c>1$. This resistance to cooperation is remarkable
because for $\delta>1$ full cooperation is a stable equilibrium of the dynamics,
and agrees with the dynamical analysis that shows that full cooperation is only
reached if the initial fraction of cooperators is already large.
When crossing $\delta_c$ cooperation suddenly invades. At that point, if we
decrease the reward again, full cooperation persists down to $\delta=1$. A
slight decrease below this point produces an abrupt spread of defection in
the population, which can even be completely invaded if $\zeta\le 1/n$. This
hysteresis is typical of critical phenomena, and it is very striking to find
it in a model like this, where na\"ive intuition says that the more one rewards
cooperation, the more cooperators should appear.
The general, most important conclusion that can be drawn from
this picture is that the effects of rewarding cooperation are neither trivial
nor as straightforward
as might be intuitively expected, and demand a more careful analysis. The origin of
this complexity lies in the dilemma that the players confront and the impossibility
to know \emph{a priori} how much reward a player can get by cooperating.

One important issue for the shared reward dilemma is where this
reward comes from. In the Introduction we have mentioned situations in Biology
that can fit the setup of the shared reward dilemma, as well as mechanisms of
direct rewarding to foster more social behavior. To name just one, companies have
realized the need of searching for mechanisms that motivate,
provide incentives or encourage
cooperative behavior among their employees in order to contribute to the effective
success of the teamwork. This context leads to another variant that we have
not considered here: the case in which the reward is detracted from the
payoff of all players. This case is particularly interesting for two reasons:
first of all, for the feedback mechanism that it implies, and secondly, because
it models a common scenario of taxation and subsequent subsidy of only certain
people. Given the complexity of the shared reward game as we have analyzed
it here, the results of this new scenario are presumed very rich.
This tax-subsidy scenario has already been explored by some
of us \citep{lugo} in a spatial evolutionary setup, but further, more 
detailed research is needed in view of the present findings. This issue
will be the subject of a forthcoming work.

In closing, we have shown that rewarding introduces a new social dilemma.
Depending on the parameters, the game casts the classical
scenarios of full defection, coexistence of cooperators and defectors,
bi-stability of full defection and full cooperation, or full cooperation, 
as well as more complex scenarios with two interior
mixed equilibria, where bi-stability between a mixed equilibrium and full
cooperation can occur.
In addition, we have seen that the cooperative response may not be
continuous on the reward, implying that promoting cooperation may require
substantial incentives.
We have shown that the classical (static) analysis of the game requires
an evolutionary (dynamic) counterpart: while in the static case the counter-intuitive
phenomenon of the decrease of the cooperation level upon increasing of the
reward may occur, this is never found dynamically; on the other hand, in
the evolutionary framework we observe that very large rewards may be
needed to establish a significant cooperation level, but once it is established,
the reward may be very much reduced without damage to the cooperative
behavior. Therefore, our general conclusion is that promoting cooperation
through a reward mechanism is far from trivial, in agreement with the non trivial
behavior found in many social contexts, and deserves careful consideration
prior to, and during, application. 

\section*{Acknowledgments}

The authors thank to Antonio Cabrales and Francisco Marhuenda,
of the Department of Economics at Universidad Carlos III de Madrid, for their
helpful suggestions on an earlier version of this paper. We also want
to thank the two anonymous referees of this paper, who have significantly
contributed to its final version.
This work is partially supported by Ministerio de Educaci\'on y Ciencia (Spain)
under grants Ingenio-MATHEMATICA, MOSAICO and NAN2004-9087-C03-03 and by Comunidad
de Madrid (Spain) under grants SIMUMAT-CM and MOSSNOHO-CM.

\appendix

\section{Characterization of symmetric Nash equilibria}
\label{sec:app-sym}

\begin{theo}
Let $\delta = \rho/n(n-1)(T-R)$ be the scaled reward of the game
and $\zeta= (T-R)/(P-S)$ the defection ratio.
Then, the following three scenarios can be found for the symmetric Nash equilibria
of the shared reward dilemma
with a number of players $n\ge 3$:
\begin{enumerate}[1.]
\item For $\zeta\geq 1/2$,
\begin{enumerate}[(i)]\setlength{\itemsep}{0pt}
\item if $\delta\leq 1/n\zeta$, the unique Nash equilibrium
is full defection ($q=0$);
\item if $1/n\zeta< \delta\leq1$, the symmetric Nash equilibrium is a continuous function
of $\delta$ which increases from $0^+$ to $1$, corresponding to the unique solution
on $(0,1]$ of
\begin{equation}
(\zeta-1)x  + 1 - \delta \zeta \frac{1-(1-x)^{n}}{x}= 0;
\label{algeq}
\end{equation}
\item if $\delta>1$ the unique Nash equilibrium
is full cooperation ($q=1$).
\end{enumerate}

\item For $1/n \leq \zeta < 1/2$,
\begin{enumerate}[(i)]
\item if $\delta\leq 1/n\zeta$ the only Nash equilibrium is full defection;
\item if $1/n\zeta< \delta < 1$ the symmetric Nash equilibrium is a continuous function
of $\delta$ which increases from $0^+$ to some limit smaller than 1,
corresponding to the unique solution on $(0,1)$ of (\ref{algeq});
\item if $\delta\ge 1$ there exists $\delta_c>1$ such that
if $\delta>\delta_c$ the unique Nash equilibrium is $q=1$,
whereas if $1\le\delta\le\delta_c$ there are two additional symmetric Nash equilibria corresponding
to the solutions $0<q_1\le q_2\leq 1$ of~(\ref{algeq}) (equality, $q_1=q_2$
holds only for $\delta=\delta_c$). The equilibria $q_1$ and $q_2$ are continuous
monotone functions of $\delta$ (increasing and decreasing respectively) and
$q_2 =1$ when $\delta=1$.
\end{enumerate}

\item For $\zeta< 1/n$,
\begin{enumerate}[(i)]
\item if $\delta <1$ the only Nash equilibrium is full defection;
\item if $1\leq \delta<1/n\zeta$ the symmetric Nash equilibria are full defection, full cooperation and
the unique solution on $(0,1]$ of (\ref{algeq}), which is a continuous function of $\delta$ which decreases from $1$ to some limit greater than 0;
\item if $\delta\ge 1/n\zeta$  there exists $\delta_c>1/n\zeta$ such that
if $\delta>\delta_c$ the unique Nash equilibrium is  $q=1$,
whereas if $1\le\delta\le\delta_c$ there are two additional symmetric Nash equilibria corresponding to
the solutions $0\leq q_1\le q_2< 1$ of~(\ref{algeq}) (equality, $q_1=q_2$ holds
only for $\delta=\delta_c$). The equilibria $q_1$ and $q_2$ are continuous monotone
functions of $\delta$ (increasing and decreasing respectively) and $q_1 =0$ when
$\delta=1/n\zeta$.
\end{enumerate}
\end{enumerate}
An upper bound for $\delta_c$ is given by
\begin{equation}
\delta_c\le \frac{1}{4\zeta} \frac{n}{(n-1)} \frac{\left(1+
\displaystyle\frac{2\zeta}{n-1}\right)^2}{
\left(1-\displaystyle\frac{n-2}{n-1}\zeta\right)}.
\end{equation}
\label{th1}
\end{theo}

\emph{Proof.} \ \
As we discussed in Section (\ref{sec:thegame}), full cooperation is a Nash equilibria iff
$\delta\geq 1$ and full defection is iff $\delta \leq 1/n\zeta$. To consider the remainder cases,
 let us define the ``loss function'' $\phi:[0,1]\rightarrow \mathbb{R}$,
\begin{equation}
\phi(x) = \frac{f_D(x) - f_C(x)}{(n-1)(P-S)} = \phi_1(x)-\delta\zeta\phi_2(x),
\end{equation}
where $\phi_1(x) = x(\zeta -1) + 1$ and 
\begin{equation}
\phi_2(x) = n\mu_{n-1}(1,x)=\begin{cases}
n, & \mbox{for $x=0$,} \\
\displaystyle{\frac{1-(1-x)^n}{x}},  & \mbox{for $0<x\le 1$.}
\end{cases}
\end{equation}
(c.f. eq. (\ref{fitness1})--(\ref{mu1})). First of all, for $\delta=0$ the only
root of the loss function is at $x=1/(1-\zeta)$, which, for any $\zeta>0$, is
outside the interval $[0,1]$. Hence $\phi(x)>0$ for all $x\in[0,1]$ and 
the only Nash equilibrium if full defection.  Let us henceforth assume $\delta>0$.
Function $\phi_2(x)$ decreases monotonically with $x$ and, for any $n>2$, is
strictly convex within the interval $[0,1]$; instead, $\phi_1(x)$ is a
straight line with nonnegative or negative slope depending on whether
$\zeta\ge 1$ or $\zeta<1$, respectively. For reasons that will be
clear in a while, we need to consider separately the cases $\zeta\ge 1$,
$\zeta<1/n$ and $1/n\le\zeta<1$.

\underline{\emph{Case $\zeta\ge 1$}}:

As $\phi_1(x)$ is nondecreasing, the loss function $\phi(x)$ monotonically
increases with $x$ and the only symmetric Nash equilibrium depends on the signs of
$\phi(0) = 1-\delta\zeta n$ and $\phi(1)= (1-\delta)\zeta$.
\begin{enumerate}[(i)]
\item If $\delta \leq 1/n\zeta$ we have $ 0\leq\phi(0)<\phi(1)$ and the 
unique Nash equilibrium is full defection. This equilibrium is strict for $\delta<1/n\zeta$.
\item If $1/n\zeta<\delta<1$ we have
$\phi(0)<0$ and $\phi(1)>0$, and the symmetric Nash equilibrium in mixed strategies 
is the solution $0< q< 1$ of (\ref{algeq}). Note that
$\phi(x)$ decreases with $\delta$, thus $q$ increases with $\delta$.
\item If $\delta\ge 1$ we have $\phi(0)<\phi(1)\le 0$ and the unique Nash equilibrium
is full cooperation, which is strict for $\delta>1$.
\end{enumerate}

\begin{figure}
\centerline{\includegraphics[width=120mm]{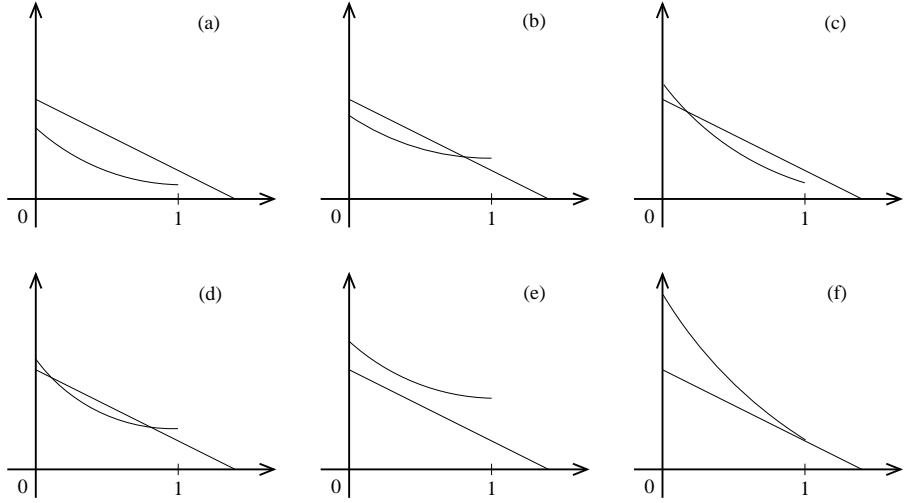}}
\caption[]{Relative situations of $\phi_1(x)$ and $\delta\zeta\phi_2(x)$ (see text).}
\label{fig:sketches}
\end{figure}

In the next two cases  $\zeta<1$ and therefore
both $\phi_1(x)$ and $\phi_2(x)$ are decreasing functions of $x$.
As $\phi_2(x)$ is convex, the situations that can occur are all
sketched in fig.~\ref{fig:sketches}.

\underline{\emph{Case $\zeta<1/n$}}:

\begin{enumerate}[(i)]
\item If $\delta < 1$ then $\phi(0)>0$ and $\phi(1) > 0$ and we have the
situation sketched in fig.~\ref{fig:sketches}(a). The only Nash
equilibrium is full defection.

\item If $1\leq \delta < 1/n\zeta$ we have $\phi(0) >  0$ and $\phi(1)\leq 0$,
so the situation is as sketched in fig.~\ref{fig:sketches}(b) and  therefore
there will be a symmetric equilibrium $0< q\leq 1$. Note that $q=1$ for $\delta=1$
and decreases as $\delta$ goes to $1/n\zeta$.

\item If $1/n\zeta\leq \delta$ then $\phi(0)\leq 0$ and $\phi(1)<0$. Thus we
will have one of the two situations plotted in
figs.~\ref{fig:sketches}(d) and \ref{fig:sketches}(e) depending on the slopes
of $\phi_1(x)$ and $\phi_2(x)$ at $x=0$ at the crossover $\delta =1/n\zeta$,
where $\phi(0)$ changes sign.
If $\phi_1'(0)>\phi_2'(0)/n$ the situation will be as illustrated in
fig.~\ref{fig:sketches}(d), and if $\phi_1'(0)\le\phi_2'(0)/n$ it
will be as in fig.~\ref{fig:sketches}(e). In the former case
there will be two Nash equilibria, $0<q_1<q_2<1$, and in the latter the only
Nash equilibrium will be $q=1$. As $\phi_1'(x)=\zeta-1$ and 
\begin{equation}
\phi_2'(x)=\frac{nx(1-x)^{n-1}-1+(1-x)^n}{x^2},
\end{equation}
we have $\phi_1'(0)=\zeta-1$ and $\phi_2'(0)=-n(n-1)/2$. The condition
$\phi_1'(0)>\phi_2'(0)/n$ reads $\zeta>(3-n)/2$, which holds for any $n\geq 3$.
We thus find two equilibria, $0\leq q_1<q_2<1$, which, upon increasing $\delta$,
approach each other ($q_1$ increases and $q_2$ decreases) up to $\delta_c$,
where they coalesce in one Nash equilibrium $q\in (0,1)$. Finally, for
$\delta>\delta_c$ the only Nash equilibrium is full cooperation.
\end{enumerate}

\underline{\emph{Case $1/n\le\zeta<1$}}:

\begin{enumerate}[(i)]
\item If $\delta<1/n\zeta$ then $\phi(0)>0$ and $\phi(1)>0$ and we have the
situation sketched in fig.~\ref{fig:sketches}(a). The only Nash
equilibrium is again $q=0$.

\item If $1/n\zeta\le\delta<1$ (this case is empty if $\zeta=1/n$)
then $\phi(0)\le 0$ and $\phi(1)>0$, and we have the situation depicted in
fig.~\ref{fig:sketches}(c). There is a unique symmetric Nash equilibrium $q\in[0,1)$
determined by (\ref{algeq}).
Also $q=0$ for $\delta = 1/n\zeta$ and increases as $\delta$ goes to 1.

\item If $\delta\ge 1$ then $\phi(0)\le 0$ and $\phi(1)\le 0$. In this
case we may have two additional equilibria if the situation of fig.~\ref{fig:sketches}(d)
occurs, or just one if either $\delta>1$ and we have the situation of
fig.~\ref{fig:sketches}(e),
or $\delta=1$ and the situation is like in fig.~\ref{fig:sketches}(f).
The separation between the first case and the last two cases depends on 
which scenario,  fig.~\ref{fig:sketches}(d) or fig.~\ref{fig:sketches}(f)
we have at $\delta=1$. This, in turn, depends on the slopes of $\phi_1(x)$
and $\phi_2(x)$ at $x=1$ when $\delta=1$: if $\phi_1'(1)<\zeta\phi_2'(1)$ then
we will have fig.~\ref{fig:sketches}(d), and if $\phi_1'(1)\ge\zeta\phi_2'(1)$ 
we will have fig.~\ref{fig:sketches}(f). The former is equivalent to
$\zeta<1/2$, the latter to $\zeta\ge 1/2$. So if $\zeta\ge 1/2$ the only 
Nash equilibrium is $q=1$, whereas if $\zeta<1/2$ there will be, for $1\le\delta
<\delta_c$, two equilibria, $0<q_1<q_2\le 1$, which coalesce in a single
one at $\delta=\delta_c$. For $\delta>\delta_c$ the only Nash equilibrium is
$q=1$.
\end{enumerate}

The limiting value $\delta_c$ can be determined as the value of $\delta$ at
which the curve $\phi_1(x)$ is tangent to $\delta_c\zeta\phi_2(x)$ at a point
$x_c\in(0,1)$. At this point the two equations
\begin{equation}
\phi_1(x_c)=\delta_c\zeta\phi_2(x_c), \qquad
\phi_1'(x_c)=\delta_c\zeta\phi_2'(x_c),
\end{equation}
hold simultaneously. These two equations can be combined to yield
\begin{eqnarray}
&& \delta_c\zeta(1-x)^n = x_c^2(1-\zeta)-x_c+\delta_c\zeta, \\
&& [(n-1)-(n-2)\zeta]x_c^2-(n-1+2\zeta)x_c+\delta_c\zeta n=0.
\end{eqnarray}
For $x_c$ to exist it is necessary that the second equation has a solution.
The condition for this to happen is
\begin{equation}
(n-1+2\zeta)^2-4[(n-1)-(n-2)\zeta]\delta_c\zeta n\ge 0.
\end{equation}
Since $\zeta<1/2$ then $(n-1)-(n-2)\zeta>0$, so the above equation holds
provided
\begin{equation}
\delta_c\le\frac{(n-1+2\zeta)^2}{4[(n-1)-(n-2)\zeta]\zeta n}
=\frac{\left(1+\frac{2\zeta}{n-1}\right)^2}{4\zeta
\left(1-\frac{n-2}{n-1}\zeta\right)}\left(\frac{n-1}{n}\right).
\end{equation}
This expresses an upper bound for $\delta_c$. $\blacksquare$

\bigskip
\begin{coro}
Consider a sequence $\{\rho_n\}$ of rewards such that $\rho_n\to\infty$
as $n\to\infty$ in such a way that
\begin{equation}
\delta=\lim_{n\to\infty}\frac{\rho_n}{n^2(T-R)},
\end{equation}
with $0\le\delta<\infty$. Let us define $\delta_\zeta=1/4\zeta(1-\zeta)$.
Then, in the limit $n\to\infty$, the Nash equilibria
of the shared reward dilemma are
\begin{enumerate}[(i)]
\item full defection if $\delta=0$;
\item a unique equilibrium in mixed strategies 
\begin{equation}
\label{equilibrio}
q = \frac{1-\sqrt{1-\delta/\delta_\zeta}}{2(1-\zeta)}
\end{equation}
if $0<\delta<1$;
\item full cooperation and two equilibria in mixed strategies, $0<q_1\le q_2<1$, where $q_1$
is given by (\ref{equilibrio}) and
\begin{equation}
q_2 = \frac{1+\sqrt{1-\delta/\delta_\zeta}}{2(1-\zeta)},
\label{q2}
\end{equation}
if  $1< \delta\leq\delta_\zeta$  and $\zeta<1/2$ (equality $q_1=q_2=1/2(1-\zeta)$
only holds if $\delta=\delta_\zeta$), and
\item full cooperation otherwise.
\end{enumerate}
\label{cor1}
\end{coro}
\emph{Proof.}~\ref{cor1}. 
As $n\to\infty$ only two of the three cases of Theorem~\ref{th1} remain,
corresponding now to $\zeta\ge 1/2$ and $0\le\zeta<1/2$. Besides,
eq.~(\ref{algeq}) becomes the quadratic equation
\begin{equation}
(\zeta-1)x^2+x-\delta\zeta=0,
\end{equation}
whose two solutions are
\begin{equation}
q_1=\frac{1-\sqrt{1-4\zeta(1-\zeta)\delta}}{2(1-\zeta)}, \qquad
q_2=\frac{1+\sqrt{1-4\zeta(1-\zeta)\delta}}{2(1-\zeta)}.
\end{equation}
Both are real whenever $0\le\delta\le\delta_{\zeta}=1/4\zeta(1-\zeta)$.
On the other hand, $q_1$ monotonically increases with $\delta$. If
$\zeta\ge 1/2$, $q_1$ runs from $0$ to $1$ as $\delta$ moves from
$0$ to $1$; if $\zeta<1/2$, $q_1$ goes from $0$ to $1/2(1-\zeta)$ as $\delta$
goes from $0$ to $\delta_{\zeta}$.
As for $q_2$, the condition for it to be within the interval $[0,1]$
is $\zeta\le 1/2$ and $1\le\delta\le\delta_{\zeta}$. When $\zeta=1/2$
and $\delta=1$ then $q_2=q_1=1$. When $\zeta<1/2$ then $q_2$ provides
a second solution, monotonically decreasing from $1$ down to $1/2(1-\zeta)$
as $\delta$ runs from $1$ to $\delta_{\zeta}$, where it coalesces with
$q_1$.

Finally, for $\delta>\delta_{\zeta}$ we have
\begin{equation}
(\zeta-1)x^2+x-\delta\zeta>0,
\end{equation}
so the only Nash equilibrium is full cooperation. $\blacksquare$

\section{Characterization of asymmetric Nash equilibria}
\label{sec:app-asym}

\begin{theo}
Let $\delta = \rho/n(n-1)(T-R)$ be the scaled reward of the game
and $\zeta= (T-R)/(P-S)$ the defection ratio. Let
\begin{equation}
\delta_k=k\,\frac{n-1+(k-1)(\zeta-1)}{n(n-1)\zeta}, \qquad
k=1,2,\dots,n-1.
\label{steps}
\end{equation}
Then a configuration with
$1\le k\le n-1$ cooperators and $n-k$ defectors will be a Nash equilibrium
in pure strategies of the shared reward dilemma if and
only if $\delta_k\le\delta\le\delta_{k+1}$ and, when $\zeta<1/2$,
$k\le(n-1)/2(1-\zeta)$.
\label{th2}
\end{theo}

\emph{Proof.} \ \
According to (\ref{pagototal}), in
a configuration with $k$ cooperators and $n-k$ defectors the payoff
of a cooperator is
\begin{equation}
\mathcal{P}_C(k)=(k-1)R+(n-k)S+\frac{\rho}{k}
\end{equation}
and of a defector
\begin{equation}
\mathcal{P}_D(k)=kT+(n-1-k)P.
\end{equation}
For such a configuration to be a Nash equilibrium in pure
strategies two requirements must be met: (i) a cooperator cannot get higher
payoff by defecting, and (ii) a defector cannot get a higher payoff by
cooperating. Condition (i) amounts to saying that
$\mathcal{P}_C(k)-\mathcal{P}_D(k-1)\ge 0$, i.e.
\begin{equation}
(k-1)(T-R)+(n-k)(P-S)-\frac{\rho}{k}\le 0,
\end{equation}
and condition (ii) amounts to saying that
$\mathcal{P}_D(k)-\mathcal{P}_C(k+1)\ge 0$, i.e.
\begin{equation}
k(T-R)+(n-1-k)(P-S)-\frac{\rho}{k+1}\le 0.
\end{equation}
By defining the parabola
\begin{equation}
\psi(x)=x^2(\zeta-1)+(n-\zeta)x-\Delta,
\label{eq:parabola}
\end{equation}
where $\Delta=\rho/(P-S)=n(n-1)\delta\zeta$,
and taking into account that $P-S>0$, the two conditions above can be
rewritten
\begin{equation}
\psi(k)\le 0, \qquad \psi(k+1)\ge 0.
\label{eq:psis}
\end{equation}
In other words, an asymmetric Nash equilibrium in pure strategies
exists if and only if there exists $k=1,2,\dots,n-1$ such that
(\ref{eq:psis}) holds.

The two roots of the parabola (\ref{eq:parabola}) are
\begin{equation}
x_{\pm}=\frac{-(n-\zeta)\pm\sqrt{(n-\zeta)^2+4\Delta(\zeta-1)}}{2(\zeta-1)},
\end{equation}
so for the discussion to follow we should treat separately the cases
$\zeta>1$, $\zeta=1$ and $\zeta<1$.

\underline{\emph{Case $\zeta>1$}.}
In this case the parabola is convex, both roots are real and $x_-<0$
and $x_+>0$. So there will be an asymmetric Nash equilibrium in pure strategies with
$k$ cooperators if and only if $k\le x_+\le k+1$, i.e.
\begin{equation}
2k(\zeta-1)\le\sqrt{(n-\zeta)^2+4\Delta(\zeta-1)}-(n-\zeta)\le 2(k+1)(\zeta-1)
\end{equation}
or equivalently
\begin{equation}
(2k-1)\zeta+n-2k\le\sqrt{(n-\zeta)^2+4\Delta(\zeta-1)}\le (2k+1)\zeta+n-2(k+1).
\label{eq:ineqA}
\end{equation}
As $\zeta>1$ we have $(2k-1)\zeta+n-2k>n-1>0$, so
all three terms in (\ref{eq:ineqA}) are positive numbers and can be
squared to obtain, after simplifying,
\begin{equation}
k\big[n-k+(k-1)\zeta\big]\le \Delta\le (k+1)(n-k-1+k\zeta).
\end{equation}
Given that $\Delta=n(n-1)\delta\zeta$, these inequalities can be rewritten
\begin{equation}
\delta_k\le\delta\le\delta_{k+1}, \qquad
\delta_k\equiv k\,\frac{n-1+(k-1)(\zeta-1)}{n(n-1)\zeta}.
\label{eq:existence}
\end{equation}
Notice that if $\zeta>1$ then $\{\delta_k\}$ forms an increasing sequence and that
$\delta_1=1/n\zeta$ and $\delta_n=1$.

\underline{\emph{Case $\zeta=1$}.}
In this case only the root $x_0=\Delta/(n-1)=n\delta$ exists,
thus the condition $k\le x_0\le k+1$ is equivalent to (\ref{eq:existence}), where, of course,
$\delta_k=k/n$.

\underline{\emph{Case $\zeta<1$}.}
The parabola (\ref{eq:parabola}) is now concave and the roots
can be rewritten
\begin{equation}
x_{\pm}=\frac{(n-\zeta)\mp\sqrt{(n-\zeta)^2-4\Delta(1-\zeta)}}{2(1-\zeta)}.
\end{equation}
For them to be real we must have
\begin{equation}
(n-\zeta)^2-4\Delta(1-\zeta)\ge 0.
\label{eq:upperbound}
\end{equation}
Suppose this inequality holds; then we have
$x_{\pm}>0$ and $x_+<x_-$. For an asymmetric Nash equilibrium
with $k$ cooperators to exist we must have $k\le x_+\le k+1\le x_-$.

The inequalities $x_+\le k+1\le x_-$ are equivalent to
\begin{equation}
|n-2k+(2k-1)\zeta|\le\sqrt{(n-\zeta)^2-4\Delta(1-\zeta)}.
\end{equation}
Squaring again this expression boils down to $\delta\le\delta_{k+1}$.
The inequality $k\le x_+$ can be rewritten
\begin{equation}
\sqrt{(n-\zeta)^2-4\Delta(1-\zeta)}\le n-2k+(2k-1)\zeta.
\end{equation}
No value of $\Delta$ satisfies this inequality unless the right-hand-side is nonnegative;
in other words, unless
\begin{equation}
k\le\frac{n-\zeta}{2(1-\zeta)}.
\label{eq:ineq2}
\end{equation}
Assuming (\ref{eq:ineq2}) holds we can square and simplify once more
to get $\delta\ge\delta_k$.

But there is one last remark to make: $\delta_k\le\delta\le\delta_{k+1}$ is empty unless
$\delta_k\le\delta_{k+1}$. If $\zeta\ge 1$ then $\delta_k$ is an increasing sequence,
but for $\zeta<1$ this is no longer true, and the constraint $\delta_k\le\delta_{k+1}$
implies
\begin{equation}
k\le\frac{n-1}{2(1-\zeta)},
\end{equation}
which is more restrictive than (\ref{eq:ineq2}). Notice that this only
constraints the value of $k$ provided $\zeta<1/2$.

Finally, one can check that (\ref{eq:upperbound}) holds for any $\delta_k$ because
\begin{equation}
(n-\zeta)^2-4n(n-1)\zeta(1-\zeta)\delta_k=\big[(2k-1)(1-\zeta)-n+1\big]^2\ge 0.
\quad \blacksquare
\end{equation}



\begin{thebibliography}{}


\harvarditem{Anderson and Franks}{2001}{anderson} Anderson, C., Franks,
	N.\ R., 2001. Teams in animal societies. Behav.\ Ecol.\ 12, 534--540.
\harvarditem{Axelrod}{1984}{axelrod:1984} Axelrod, R., 1984. The Evolution
	of Cooperation. Penguin, London.
\harvarditem{Axelrod and Hamilton}{1981}{axelrod:1981} Axelrod, R., and Hamilton,
        W.\ D., 1981. The evolution of co-operation. Science 211, 1390--1396.
\harvarditem{Chao and Strawderman}{1972}{chao:1972} Chao, M.\ T.,
	Strawderman, W.\ E., 1972. Negative moments of positive random variables.
	J.\ Am.\ Stat.\ Soc.\ 67, 429--431.
\harvarditem{Cressman}{2003}{Cr03} Cressman, R., 2003. Evolutionary dynamics
	and extensive form games. MIT Press, Cambridge, Massachusetts.
\harvarditem{Doebeli and Hauert}{2005}{doebeli-hauert:2005} Doebeli, M.,
	Hauert, C., 2005. Models of cooperation based on the Prisoner's Dilemma
	and the Snowdrift game. Ecol.\ Lett.\ 8, 748--766.
\harvarditem{Egu\'iluz \emph{et al.}}{2005}{maxy:2005} Egu\'iluz, V.,
	Zimmermann, M., Cela-Conde, M.\ G., San Miguel, M., 2005. Cooperation
	and the emergence of role differentiation in the dynamics of social
	networks. Am. \ J.\ Soc.\ 110, 977--1008.
\harvarditem{Gintis}{2000}{Gintis:2000} Gintis, H., 2000. Game theory evolving.
	Princeton University Press, Princeton.
\harvarditem{Hauert \emph{et al.}}{2006}{hauert:2006} Hauert, C., Michor, F.,
	Nowak, M., Doebeli, M., 2006. Synergy and discounting of cooperation
	in social dilemmas. J.\ Theor.\ Biol.\ 239, 195--202 
\harvarditem{Hauert and Szab\'o}{2003}{hauert:2003} Hauert, C., Szab\'o, G., 2003.
        Prisoner's dilemma and public goods games in different geometries:
        compulsory versus voluntary participation. Complexity 8, 31--38.
\harvarditem{Hamilton}{1964a}{hamiltona} Hamilton, W.D., 1964a. The genetical
        evolution of social behaviour I. J.\ Theor.\ Biol.\ 7, 1--16.
\harvarditem{Hamilton}{1964b}{hamiltonb} Hamilton, W.D., 1964b. The genetical
        evolution of social behaviour II. J.\ Theor.\ Biol.\ 7, 17--52.
\harvarditem{Hardin}{1968}{hardin:1968} Hardin, G., 1968. The Tragedy of the Commons.
        Science, 162, 1243--1248.
\harvarditem{Hauert \emph{et al.}}{2007}{hauert:2007} Hauert, C., Traulsen, A.,
        Brandt, H., Nowak, M.\
	A., and Sigmund, K., 2007. Via freedom to coercion: the emergence of
	costly punishment. Science 316, 1905--1907.
\harvarditem{Hofbauer and Sigmund}{1998}{hofbauer-sigmund:1998} Hofbauer, J.,
	Sigmund, K, 1998. Evolutionary Games and Population Dynamics. Cambridge
	University Press, Cambridge.
\harvarditem{Iribarren and Moro}{2007}{iribarren} Iribarren, J.\ L., Moro, E.,
	2007. Information diffusion epidemics in social networks.
	http://arxiv.org/pdf/0706.0641.
\harvarditem{Jim\'enez \emph{et al.}}{2007}{auto1} Jim\'enez, R., Lugo, H.,
	Cuesta, J.\ A., S\'anchez, A., 2007. Emergence and resilience of
	cooperation in the Spatial Prisoner's Dilemma via a reward mechanism.
	Working paper.
\harvarditem{Kiers \emph{et al.}}{2003}{kiers:2003} Kiers, E.\ T., Rousseau,
	R.\ A., West, S.\ A., Deniso, R.\ F., 2003. Host sanctions and the
	legumerhizobium mutualism. Nature 425, 78--81.
\harvarditem{Licht}{1999}{licht:1999} Licht, A.\ N., 1999. Games commissions
	play: 2x2 fames of international securities regulation. Yale J.\ Int.\
	Law 24, 61--125.
\harvarditem{Lugo and Jim\'enez}{2006}{lugo} Lugo, H., Jim\'enez, R., 2006.
	Incentives to Cooperate in Network Formation. Computational Economics
	28, 15--27.
\harvarditem{Maynard-Smith and G. Price}{1973}{maynard-price:1973} Maynard-Smith,
	J., Price, G., 1973. The logic of animal conflict. Nature \ 246, 15--18.
\harvarditem{Moro}{2004}{moro} Moro, E., 2004. The Minority Game: an introductory
	guide. Advances in Condensed Matter and Statistical Physics. Korutcheva,
        E., and Cuerno, R., eds. Nova Science Publishers, 263--286.
\harvarditem{Nowak}{2006}{nowak:2006} Nowak, M.\ A., 2006. Five rules for the
	evolution of cooperation. Science, 314, 1560--1563.
\harvarditem{Nowak and May}{1992}{nowak-may:1992} Nowak, M.\ A., May, R.\ M., 1992.
	Evolutionary games and spatial chaos. Nature 415, 424--426.
\harvarditem{Nowak and Sigmund}{1998}{nowaksigmund:1998} Nowak, M.\ A., Sigmund, K.,
        1998. Evolution of indirect reciprocity by image scoring. Nature 393,
	573--577.
\harvarditem{Nowak and Sigmund}{2000}{nowak:2000} Nowak, M.\ A., Sigmund, K., 2000.
	Games on grids. The Geometry of Ecological Interactions. Dieckmann, U.,
	Law, R., and Metz, J.\ A.\ J., eds. Cambridge University Press, 135--150.
\harvarditem{Packer and Ruttan}{1988}{packer} Packer, C., Ruttan, L., 1988. The
	evolution of cooperative hunting. Am.\ Nat.\ 132, 159--198.
\harvarditem{Pennisi}{2005}{pennisi} Pennisi, E., 2005. How did cooperative behavior
        evolve? Science 309, 93.
\harvarditem{Sigmund \emph{et al.}}{2001}{sigmund:2001} Sigmund, K., Hauert, C.,
	Nowak, M.\ A., 2001. Reward and punishment. Proc.\ Nat.\ Acad.\ Sci.\ 98,
	10757--10762.
\harvarditem{Skyrms}{2003}{skyrms:2003} Skyrms, B., 2003. The stag hunt and
	Evolution of Social Structure. Cambridge University Press, Cambridge.
\harvarditem{Sugden}{1986}{sugden:1986} Sugden, R., 1986. The economics of rights,
	co-operation and welfare. Basil Blackwell, Oxford.

\end{thebibliography}
\end{document}